\documentclass[namedreferences,optionalrh]{solarphysics}
\usepackage[hyperref,optionalrh,solaromanenum]{spr-sola-addons} % For Solar Physics 

\usepackage[svgnames,dvipsnames]{xcolor}
\usepackage{xparse}
\usepackage{bm}
\usepackage{latexsym}
\usepackage[
capitalize,
noabbrev
]{cleveref}
\usepackage{tikz}
\usepackage[normalem]{ulem}

\newcommand{\BB}{\ensuremath{\bm{B}}}

\NewDocumentCommand{\curl}{o}{\ensuremath{\nabla \times #1}}
\NewDocumentCommand{\grad}{o}{\ensuremath{\nabla #1}}
\NewDocumentCommand{\diverg}{o}{\ensuremath{\nabla \cdot #1}}
\NewDocumentCommand{\years}{o}{\ensuremath{
\IfNoValueF{#1}{#1 \,}
\mathrm{years}
}}

\NewDocumentCommand{\days}{o}{\ensuremath{
\IfNoValueF{#1}{#1 \,}
\mathrm{days}
}}

\NewDocumentCommand{\months}{o}{\ensuremath{
\IfNoValueF{#1}{#1 \,}
\mathrm{months}
}}

\NewDocumentCommand{\s}{o}{\ensuremath{
\IfNoValueF{#1}{#1 \,}
\mathrm{s}
}}

\NewDocumentCommand{\nT}{o}{\ensuremath{
\IfNoValueF{#1}{#1 \,}
\mathrm{nT}
}}

\NewDocumentCommand{\km}{o}{\ensuremath{
\IfNoValueF{#1}{#1 \,}
\mathrm{km}
}}

\NewDocumentCommand{\au}{o}{\ensuremath{
\IfNoValueF{#1}{#1 \,}
\mathrm{AU}
}}

\NewDocumentCommand{\temp}{o}{\ensuremath{
\IfNoValueF{#1}{#1 \times}
\mathrm{10^5 \; K}
}}

\NewDocumentCommand{\cc}{o}{\ensuremath{
\IfNoValueF{#1}{#1 \;}
\mathrm{cm}^{-3}
}}

\NewDocumentCommand{\pct}{o}{\ensuremath{
\IfNoValueF{#1}{#1 \;}
\%
}}

\NewDocumentCommand{\Rs}{o}{\ensuremath{
\IfNoValueF{#1}{#1 \;}
\mathrm{R_S}
}}

\NewDocumentCommand{\kms}{o}{\ensuremath{
\IfNoValueF{#1}{#1 \;}
\mathrm{km \, s^{-1}}
}}

\NewDocumentCommand{\mWcc}{o}{\ensuremath{
\IfNoValueF{#1}{#1 \;}
\mathrm{mW \cc}
}}
\newcommand{\he}{\ensuremath{\mathrm{He}}}

\newcommand{\vsw}{\ensuremath{v_\sw}}

\NewDocumentCommand{\grate}{o o}{\ensuremath{
\gamma\IfNoValueF{#1}{/\Omega_{#1}}
\IfNoValueF{#2}{= 10^{{#2}}}
}}

\NewDocumentCommand{\gmax}{o}{\ensuremath{
\gamma_\mathrm{max}\IfNoValueF{#1}{/\Omega_{#1}}
}}

\NewDocumentCommand{\kvec}{o}{\ensuremath{
\vec{k} \rho\IfNoValueF{#1}{{_{#1}}}
}}

\NewDocumentCommand{\kpar}{o}{\ensuremath{
{k_\parallel} \rho\IfNoValueF{#1}{{_{#1}}}
}}

\NewDocumentCommand{\kper}{o}{\ensuremath{
{k_\perp} \rho\IfNoValueF{#1}{{_{#1}}}
}}

\NewDocumentCommand{\ani}{s o}{\ensuremath{
R\IfNoValueF{#2}{_{#2}}
\IfBooleanT{#1}{\, [\perp\!/\!\parallel]}
}}

\NewDocumentCommand{\Trat}{s m m o}{\ensuremath{
T_{\IfNoValueF{#4}{{#4};}#2}/T_{\IfNoValueF{#4}{{#4};}#3}
 \IfBooleanT{#1}{\, [\#]}
}}

\NewDocumentCommand{\pbeta}{s o}{\ensuremath{
\beta\IfNoValueF{#2}{_{#2}}
 \IfBooleanT{#1}{\, [\#]}
}}

\NewDocumentCommand{\pbetaR}{o}{\ensuremath{
(\pbeta[\parallel
\IfNoValueF{#1}{;#1}], \ani[#1])
}}

\NewDocumentCommand{\dv}{o}{\ensuremath{\Delta v\IfNoValueF{#1}{_{#1}}}}
\NewDocumentCommand{\ca}{o}{\ensuremath{C_{A\IfNoValueF{#1}{;#1}}}}
\NewDocumentCommand{\dvca}{o o}{\ensuremath{\dv[#1]/\ca[#2]}}

\NewDocumentCommand{\nuc}{o}{\ensuremath{\nu_{c\IfNoValueF{#1}{;#1}}}}
\NewDocumentCommand{\Nc}{o}{\ensuremath{N_{c\IfNoValueF{#1}{;#1}}}}
\NewDocumentCommand{\Ac}{o}{\ensuremath{A_{c\IfNoValueF{#1}{;#1}}}}

\NewDocumentCommand{\tauEXP}{o}{\ensuremath{
\tau_{\mathrm{exp}\IfNoValueF{#1}{;#1}
}}}

\NewDocumentCommand{\tauCC}{o}{\ensuremath{
\tau_{\mathrm{C}\IfNoValueF{#1}{;#1}
}}}

\newcommand{\Lalpha}{\ensuremath{\mathrm{L}\alpha}}

\newcommand{\sw}{\ensuremath{\mathrm{sw}}}

\NewDocumentCommand{\qpar}{o}{\ensuremath{
q_{\parallel
\IfNoValueF{#1}{;#1}
}}}

\NewDocumentCommand{\edv}{o}{\ensuremath{
\tilde{E}_{\dv[#1]
}}}

\NewDocumentCommand{\se}{o}{\ensuremath{
S{\IfNoValueF{#1}{_{#1}}}
}}

\NewDocumentCommand{\ab}{o}{\ensuremath{
A{\IfNoValueF{#1}{_{#1}}}
}}
\newcommand{\ahe}{\ab[\he]}

\NewDocumentCommand{\xcorr}{o}{\ensuremath{
\rho
\IfNoValueF{#1}{(#1)}
}}

\NewDocumentCommand{\xhel}{o}{\ensuremath{
\sigma_{c
\IfNoValueF{#1}{,#1}
}
}}

\newcommand{\nst}[1]{#1\textsuperscript{st}}
\newcommand{\nd}[1]{#1\textsuperscript{nd}}

\newcommand{\nth}[1]{#1\textsuperscript{th}}

\newcommand{\degree}{\ensuremath{^\circ}}

\definecolor{q}{HTML}{228B22}
\definecolor{wc}{HTML}{FF8C00}
\definecolor{dnc}{HTML}{FF00FF}
\definecolor{todo}{HTML}{e13748}
\definecolor{ben}{HTML}{e13748}
\definecolor{bob}{HTML}{0080FF}

\NewDocumentCommand{\question}{s o m}{\IfBooleanF{#1}{\textcolor{q}{\textbf{Q}\IfNoValueF{#2}{ (#2)}: \textit{#3}}}}
\NewDocumentCommand{\answer}{s o m}{\IfBooleanF{#1}{\textcolor{q}{\textbf{A}\IfNoValueF{#2}{ (#2)}: \textit{#3}}}}

\NewDocumentCommand{\wc}{s m}{\IfBooleanTF{#1}{#2}{\textcolor{wc}{\textbf{WC:} \textit{#2}}}}
\NewDocumentCommand{\ws}{s m}{\IfBooleanTF{#1}{#2}{\textcolor{wc}{\textbf{WS:} \textit{#2}}}}

\NewDocumentCommand{\delete}{s m}{\IfBooleanF{#1}{\textcolor{todo}{\textbf{Delete:} \textit{#2}}}}
\NewDocumentCommand{\todo}{s o m}{\IfBooleanF{#1}{\textcolor{todo}{\textbf{TODO}\IfNoValueF{#2}{ (#2)}: \textit{#3}}}}
\NewDocumentCommand{\verify}{s o m}{\IfBooleanTF{#1}{#3}{\textcolor{todo}{\textbf{VERIFY}\IfNoValueF{#2}{ (#2)}: \textit{#3}}}}
\NewDocumentCommand{\goal}{s o m}{\IfBooleanTF{#1}{#3}{\textcolor{todo}{\textbf{GOAL}\IfNoValueF{#2}{ (#2)}: \textit{#3}}}}

\NewDocumentCommand{\move}{s o m}{\textcolor{dnc}{\textbf{\IfBooleanTF{#1}{Duplicate}{Move}}\IfNoValueF{#2}{ (#2)}: \textit{#3}}}
\NewDocumentCommand{\dupe}{o m}{\move*[#1]{#2}}
\NewDocumentCommand{\intro}{s m}{\IfBooleanTF{#1}{\dupe[Intro]{#2}}{\move[DnC]{#2}}}
\NewDocumentCommand{\dnc}{s m}{\IfBooleanTF{#1}{\dupe[DnC]{#2}}{\move[DnC]{#2}}}
\NewDocumentCommand{\fw}{s m}{\IfBooleanTF{#1}{\dupe[Future Work]{#2}}{\move[DnC]{#2}}}

\DeclareRobustCommand{\PlusEmpty}{\ensuremath{\mathord{\begin{tikzpicture}[line width=0.3ex, x=1.25ex, y=1.25ex]
\draw (0.5, 0.5) -- (0,0.5) -- (0, 1) -- (0.5, 1) -- (0.5, 1.5) -- (1, 1.5) -- (1, 1) -- (1.5, 1) -- (1.5, 0.5) -- (1, 0.5) -- (1, 0) -- (0.5, 0) -- cycle;
\end{tikzpicture}}}}

\DeclareRobustCommand{\SolidBand}{\ensuremath{\mathord{\begin{tikzpicture}[line width=1.25ex, x=1.25ex, y=1.25ex, yshift=5ex]
\draw (0,0.5) -- (1.5,0.5);
\draw[opacity=0, line width=0.1ex] (0,0) -- (1.5,0);
\end{tikzpicture}}}}

\DeclareRobustCommand{\VertLine}{\ensuremath{\mathord{\begin{tikzpicture}[line width=0.3ex, x=1.25ex, y=1.25ex, yshift=5ex]
\draw[opacity=0, line width=0.1ex] (0,1.25) -- (0.15,1.25);
\draw (0,0) -- (0,1.25);
\end{tikzpicture}}}}

\DeclareRobustCommand{\SolidLine}{\ensuremath{\mathord{\begin{tikzpicture}[line width=0.3ex, x=1.25ex, y=1.25ex, yshift=5ex]
\draw (0,0.5) -- (1,0.5);
\draw[opacity=0] (0,0) -- (1,0);
\end{tikzpicture}}}}

\DeclareRobustCommand{\DashedLine}{\ensuremath{\mathord{\begin{tikzpicture}[line width=0.3ex, x=1.25ex, y=1.25ex, yshift=5ex]
\draw (0,0.5) -- (0.75,0.5);
\draw (1.0,0.5) -- (1.75,0.5);
\draw[opacity=0] (0,0) -- (1.25,0);
\end{tikzpicture}}}}

\DeclareRobustCommand{\DashDotLine}{\ensuremath{\mathord{\begin{tikzpicture}[line width=0.3ex, x=1.25ex, y=1.25ex, yshift=5ex]
\draw (0,0.5) -- (0.7,0.5);
\draw (0.9,0.5) -- (1.2,0.5);
\draw[opacity=0] (0,0) -- (1.2,0);
\end{tikzpicture}}}}

\DeclareRobustCommand{\DotDotDotLine}{\ensuremath{\mathord{\begin{tikzpicture}[line width=0.3ex, x=1.25ex, y=1.25ex, yshift=5ex]
\draw (0,0.5) -- (0.3,0.5);
\draw (0.6,0.5) -- (0.9,0.5);
\draw (1.2,0.5) -- (1.5,0.5);
\draw[opacity=0] (0,0) -- (1.5,0);
\end{tikzpicture}}}}

\DeclareRobustCommand{\DashDotDotDotLine}{\ensuremath{\mathord{\begin{tikzpicture}[line width=0.3ex, x=1.25ex, y=1.25ex, yshift=5ex]
\draw (0,0.5) -- (1,0.5);
\draw (1.3,0.5) -- (1.6,0.5);
\draw (1.9,0.5) -- (2.2,0.5);
\draw (2.5,0.5) -- (2.8,0.5);
\draw[opacity=0] (0,0) -- (2.8,0);
\end{tikzpicture}}}}

\DeclareRobustCommand{\Circle}{\ensuremath{\mathord{\begin{tikzpicture}[line width=0.3ex, x=1.25ex, y=1.25ex, yshift=5ex]
\draw circle (0.75ex);
\end{tikzpicture}}}}

\NewDocumentCommand{\sect}{o m}{Section~\ref{sec:#2}\IfNoValueF{#1}{ #1}}
\NewDocumentCommand{\fig}{o m}{\cref{fig:#2}\IfNoValueF{#1}{ #1}}
\NewDocumentCommand{\eq}{o m}{\cref{eq:#2}\IfNoValueF{#1}{ #1}}
\NewDocumentCommand{\tbl}{o m}{\cref{tbl:#2}\IfNoValueF{#1}{ #1}}

\usepackage{graphicx}                    % For eps figures, newer & more powerfull
\usepackage[counterclockwise]{rotating}
\usepackage{caption}
\usepackage[rightcaption]{sidecap}
\usepackage{breakurl}                         % For breaking URLs easily trough lines
\graphicspath{{./}{figures/}}
\DeclareGraphicsExtensions{.pdf,.png,.jpg,.eps,}

\newcommand{\plotSWEAbundance}{
\begin{sidewaysfigure}
\includegraphics[width=\textwidth]{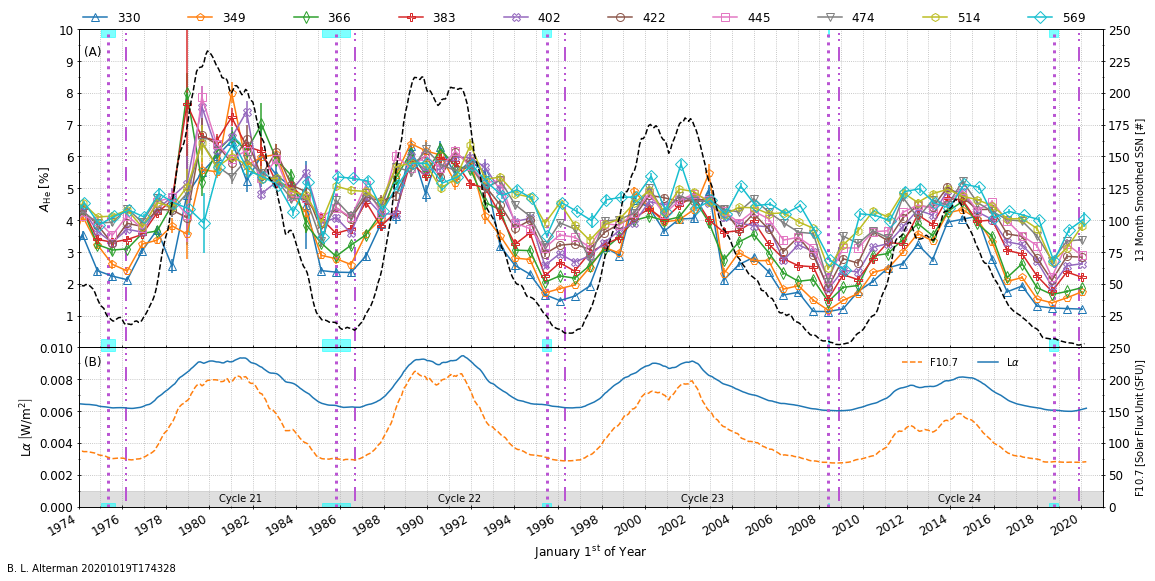}
\caption{\textbf{(A)} OMNI/Lo helium abundance (\ab[\he]) as a function of solar wind speed (\vsw) and time.
\ab[\he] is split into 10 \vsw-quantile, each indicated by a unique color and marker.
The legend at the figure's top indicates each quantile's center in \kms. 
Within each \vsw-quantile, \ab[\he] is averaged down to 250-day resolution.
Error bars indicate the standard error of the mean.
The secondary y-axis plots the 13 month smoothed sunspot number (SSN, $\bm{\DashedLine}$).
Vertical dash-dotted purple lines (\textcolor[HTML]{ba55d3}{$\DashDotDotDotLine$}) indicate Solar Cycle Minima, which conventionally indicate the start of a new solar cycle.
Vertical dotted lines (\textcolor[HTML]{ba55d3}{$\DotDotDotLine$}) indicate the \ahe\ Shutoff date averaged across \vsw-quantiles.
Blue boxes (\textcolor[HTML]{7fffff}{$\blacksquare$}) at the top and bottom of each panel indicate the standard deviation of these shutoff dates.
\textbf{(B)} Lyman-$\alpha$ (\Lalpha, left, \textcolor[HTML]{1f77b4}{$\SolidLine$}) and F10.7 cm radio emission (right, \textcolor[HTML]{ff7f0e}{$\DashedLine$}) solar activity indicators over the same period as \ab[\he].
Within one or two data points, \textbf{\ab[\he] increases across all but the one or two slowest \vsw-quantiles prior to plotted SSN Minima}.
}
\label{fig:SWE-abundance}
\end{sidewaysfigure}
}

\newcommand{\plotAhePreceedingSolarMin}{
\begin{figure}
\includegraphics[width=\textwidth]{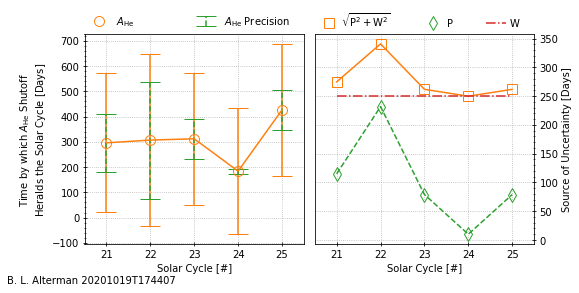}
\caption{
\textbf{(Left)} 
Orange circles (\textcolor[HTML]{ff7f0e}{$\bm{\Circle}$}) plot the time by which \ahe\ Shutoff heralds a given solar cycle.
Green error bars (\textcolor[HTML]{2ca02c}{\DashedLine}) indicate the uncertainty only due to \ahe's precision ($\mathrm{P}$).
Orange error bars (\textcolor[HTML]{ff7f0e}{\SolidLine}) show the uncertainty due to the geometric mean of the precision and the 250-day averaging window length ($\sqrt{\mathrm{P}^2 + \mathrm{W}^2}$).
\textbf{(Right)} Sources of uncertainty in each of the (\emph{Left}) panel's data points.
Orange diamonds (\textcolor[HTML]{ff7f04}{$\bm{\lozenge}$}) and green squares (\textcolor[HTML]{2ca02c}{$\bm{\square}$}) plot the uncertainties indicated in the same color that the (\emph{Left}) panel uses.
The red dash-dotted line (\textcolor[HTML]{d62728}{\DashDotLine}) indicates the 250-day averaging window ($\mathrm{W}$).
The 250-day averaging window is the dominant source of uncertainty in determining when \ahe\ Shutoff occurs.
}
\label{fig:Shutoff-dt-SolMin}
\end{figure}
}

\newcommand{\plotAheCycleLength}{
\begin{SCfigure}
\includegraphics[width=0.5\textwidth]{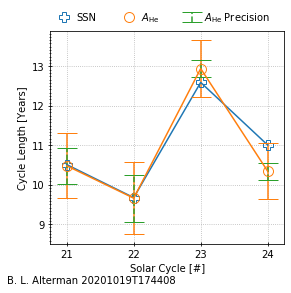}
\caption{
Blue pluses (\textcolor[HTML]{1f77b4}{$\bm{\PlusEmpty}$}) plot SSN cycle length as a function of cycle number as conventionally defined.
Orange circles (\textcolor[HTML]{ff7f0e}{$\bm{\Circle}$}) plot the time between successive \ahe\ Shutoffs as a function of the solar cycle during a given \ahe\ Shutoff interval closing a Shutoff cycle occurs.
This is $N-1$ in comparison to the cycle number a given \ahe\ Shutoff heralds (c.f.~\fig{Shutoff-dt-SolMin}).
As in \fig{Shutoff-dt-SolMin}'s (\emph{Left}) Panel, Orange error bars (\textcolor[HTML]{ff7f0e}{\SolidLine}) indicates the uncertainty due to \ahe's precision \& the 250-day averaging window.
Green error bars (\textcolor[HTML]{2ca02c}{\DashedLine}) indicate the uncertainty due to \ahe's precision alone.
}
\label{fig:Shutoff-Cycle-Length}
\end{SCfigure}
}

\newcommand{\plotAheBP}{
\begin{figure}
\includegraphics[width=\textwidth]{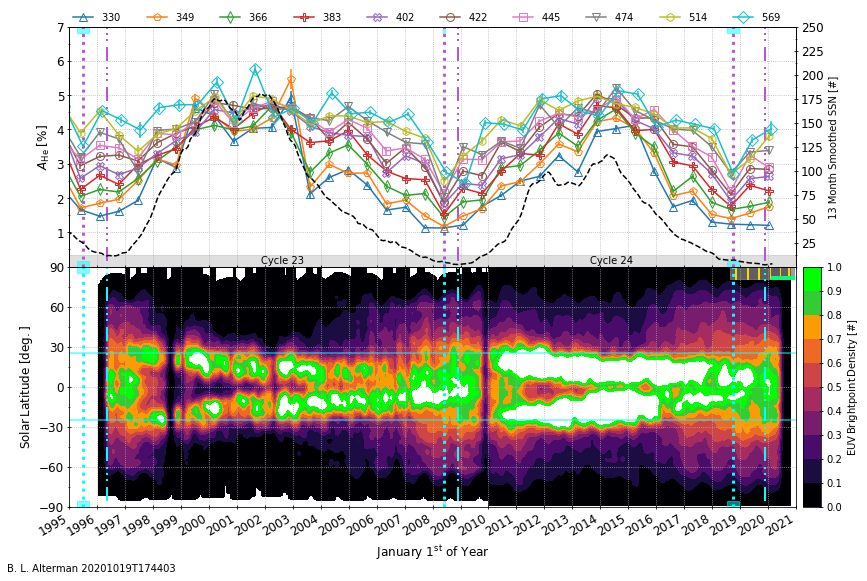}
\caption{\textbf{(Top)} Solar wind helium abundance (\ahe) as a function of solar wind speed (\vsw) and time as in \fig{SWE-abundance} over the time period covering 1995 to present day.
\textbf{(Bottom)} A filled contour plot of the combined daily SOHO/EIT (195\AA) and SDO/AIA (193\AA) EUV central meridian Brightpoint (BP) density as a function of solar latitude and time.
BPs are averaged down to 27 day cadence.
The color scale is chosen to highlight BP density $> 0.8$ and levels $> 1$ are shown in white to emphasize BPs in the range 0.8 to 1.0.
Solar Minima and \ahe\ Shutoffs are plotted in light blue to contrast with BP density.
Partially transparent blue lines indicate the latitudes $\lambda = \pm 25\degree$.
The gray band (\textcolor[HTML]{707071}{$\SolidBand$}) indicates Parker Solar Probe (PSP) $launch - \days[70]$ through the exit of its 6\textsuperscript{th} near-Sun $+ \days[70]$ and the yellow bars (\textcolor[HTML]{ffd700}{$\VertLine$}) indicate the time spent below \au[0.25].
The green bar (\textcolor[HTML]{00ff7f}{$\SolidBand$}) that starts at PSP encounter 4 indicates the time period following Solar Orbiter's (SolO) Feb 2020 launch \citep{Muller2020}.
}
\label{fig:Ahe-BP}
\end{figure}
}

\newcommand{\eqahe}{
\begin{equation} \label{eq:ahe}
\ab[\he] = 100 \times n_\mathrm{He}/n_\mathrm{H},
\end{equation}
}

\newcommand{\colheader}[1]{\multicolumn{1}{l}{#1}}

\newcommand{\MinimaTable}{
\begin{table}
\begin{tabular}{l rrrrrrrr}
\hline
\colheader{Solar} &       \colheader{\ahe\ Shutoff} & \multicolumn{3}{l}{Heralding Statistics} & \multicolumn{4}{l}{Cycle Length} \\
\colheader{Cycle} &       \colheader{Date} & \colheader{Time}  &   \colheader{Prec.}  &  \colheader{Uncert.} & \multicolumn{2}{l}{\ahe\ Shutoff} & \multicolumn{2}{l}{SSN} \\
$[\#]$ &    [YYYY-MM-DD]  & [D] &  [D]  & [D] & [250 D] & [Y] & [250 D] & [Y]\\
\hline
21           &  1975-05-10 &         296 &  115 &             275 &          15.3 &          10.5 &        15.3 &        10.5 \\
22           &  1985-10-29 &         307 &  232 &             341 &           14.1 &          9.7 &         14.1 &         9.7 \\
23           &  1995-06-24 &         312 &   79 &             262 &          18.9 &          12.9 &        18.4 &        12.6 \\
24           &  2008-05-31 &         184 &   10 &             250 &          15.1 &          10.3 &         16.1 &       11.0 \\
25           &  2018-10-01 &         426 &   79 &             262 &           --- &         --- &          --- &        --- \\
\hline
\end{tabular}
\caption{
\ahe\ Shutoff statistics averaged over \vsw-quantiles for the the indicated Solar Cycles.
Heralding Time is the time by which \ahe\ Shutoff precedes the indicated Solar Minimum.
\ahe\ Shutoff Date is the date at which Shutoff occurred.
\ahe\ Shutoff precision ($\mathrm{Prec.}$) is the standard deviation of the Shutoff Date.
\ahe\ Shutoff uncertainty $= \sqrt{\mathrm{P}^2 + \mathrm{W}^2}$ ($\mathrm{Uncert.}$) is the geometric mean of the precision and 250-day averaging window ($\mathrm{W}$).
The average Heralding Time weighted by Shutoff Precision is \days[229] and weighted by Shutoff Uncertainty is \days[302].
Cycle Length gives the time between successive \ahe\ Shutoffs for the Solar Cycle in which the \ahe\ Shutoff that closes a given \ahe\ Cycle occurs.
\ahe\ Shutoff cycle length is typically comparable to SSN cycle length.
}
\label{tbl:ahe-minima}
\end{table}
}

\newcommand{\MinFootnotes}{\footnote{\href{http://sidc.be/silso/node/167/\#NewSolarActivity}{sidc.be/silso/node/167/\#NewSolarActivity}}\textsuperscript{,}\footnote{\href{https://www.nasa.gov/press-release/solar-cycle-25-is-here-nasa-noaa-scientists-explain-what-that-means/}{nasa.gov/press-release/solar-cycle-25-is-here-nasa-noaa-scientists-explain-what-that-means}}}

\begin{document}

\begin{article}

\begin{opening}
% !TEX root = manuscript.tex
\title{Solar Wind Helium Abundance Heralds Solar Cycle Onset}
\runningauthor{Alterman et al.}
\runningtitle{Helium Heralds Solar Cycle Onset}

\author[addressref={SwRI,CLaSP},corref,email={blalterman@swri.org}
]{\inits{B.L.}\fnm{Benjamin L.}~\lnm{Alterman}\orcid{0000-0001-6673-3432}}

\author[addressref={BWXT,CLaSP},email={jckasper@umich.edu}
]{\inits{J.C.}\fnm{Justin C.}~\lnm{Kasper}\orcid{0000-0002-7077-930X}}

\author[addressref={UMBC,GSFC},email={robert.j.leamon@nasa.gov}
]{\inits{R.J.}\fnm{Robert J.}~\lnm{Leamon}\orcid{0000-0002-6811-5862}}

\author[addressref={NCAR},email={mscott@ucar.edu}
]{\inits{S.W.}\fnm{Scott W.}~\lnm{McIntosh}\orcid{0000-0002-7369-1776}}

\address[id=SwRI]{Space Science and Engineering, 
Southwest Research Institute, 
6220 Culebra Road, 
San Antonio, TX 78238, USA}
\address[id=CLaSP]{University of Michigan, 
Department of Climate \& Space Sciences \& Engineering, 
2455 Hayward St., 
Ann Arbor, MI 48109-2143, USA}
\address[id=BWXT]{BWX Technologies, Inc.
Washington, D.C.~20002, USA}
%\address[id=CfA]{Center for Astrophysics, 
%Observatory Building E, 
%60 Garden St., 
%Cambridge, MA 02138, USA}
\address[id=UMBC]{University of Maryland--Baltimore County, 
Goddard Planetary Heliophysics Institute,
Baltimore, MD 21250, USA}
\address[id=GSFC]{NASA Goddard Space Flight Center,  
Code 672,  
Greenbelt, MD 20771, USA}
\address[id=NCAR]{National Center for Atmospheric Research,  
High Altitude Observatory,  
P.O.\ Box 3000,  
Boulder, CO~80307, USA}

% !TEX root =	../manuscript.tex

\begin{abstract}

We study the solar wind helium-to-hydrogen abundance's (\ahe) relationship to solar cycle onset.
Using OMNI/Lo data, we show that \ahe\ increases prior to sunspot number (SSN) minima.
We also identify a rapid depletion and recovery in \ahe\ that occurs directly prior to cycle onset.
This \ahe\ Shutoff happens at approximately the same time across solar wind speeds (\vsw), implying that it is formed by a mechanism distinct from the one that drives \ahe's solar cycle scale variation and \vsw-dependent phase offset with respect to SSN.
The time between successive \ahe\ shutoffs is typically on the order of the corresponding solar cycle length.
Using Brightpoint (BP) measurements to provide context, we infer that this shutoff is likely related to the overlap of adjacent solar cycles and the equatorial flux cancelation of the older, extended solar cycle during Solar Minima.
%\todo{Comparing \ahe's pre-Minimum 25 shutoff with the most recent consensus prediction and the announcement of Cycle 25 demonstrates that \ahe\ can herald Solar Cycle onset before the SSN.
%This comparison also suggests that Solar Cycle 25 may have started before the SSN indicates.}
%As \ahe's rapid depletion and recovery have already occurred and \ahe\ is now increasing as it has following previous solar minima, we infer that solar cycle 25 has already begun.

\end{abstract}

\end{opening}
%-------------------------------------------------

%%%%%%%%%%%%%%%%%%%%%%%%%%%%%%%%%%%%%%%%%%%%%%%%%%%
%% Sections
% !TEX root =	../manuscript.tex

\section{Introduction} 
\label{sec:intro}

Since at least \citeyear{Schwabe1844}, the Sun's approximately 11-year solar activity cycle has been measured in the sunspot number (SSN) \citep{Schwabe1844}.
Today, many other activity indices are known to track the solar cycle.
Typically, these carry a known phase offset when measured with respect to SSN.
For example, Lyman-$\alpha$ (\Lalpha) lags SSN by 125 days \citep{Bachmann1994} and soft X-ray flux (SXR) lags SSN by 300 to 450 days \citep{Temmer2003}.

Helium is a natural byproduct of Big Bang nucleosynthesis and solar fusion \citep{Bethe1938,Bethe1939,Parker1997,Basu2008}.
It composes $\sim25\%$ of solar material by mass \citep{Basu2008,Asplund2009,Laming2015a,Basu2004} and is the most common solar element after hydrogen.
The first ionization potential (FIP) is the energy necessary to ionize a neutral atom's \nst{1} electron.
As helium has the highest FIP of any solar element, it is the last to ionize in the upper convection zone \citep{Basu2008,Laming2015a}.
Through the chromosphere and transition region, the FIP effect depletes the helium abundance \citep{Laming2015a,Rakowski2012} such that, outside of transient events like coronal mass ejections (CMEs), it drops to below 5\% by the time its is released into the solar wind \citep{Asplund2009,Laming2015a,Hirshberg1973,Neugebauer1981,Aellig2001,Kasper2007,Kasper2012,Alterman2019}.

\citet{Neugebauer1962} made the first in situ helium measurements with Mariner II.
The helium abundance is given by
\eqahe
where $n_\he$ is the helium number density and $n_\mathrm{H}$ is the hydrogen number density.
In the intervening 58 years, multiple authors have shown that \ahe\ tracks the solar cycle \citep{Ogilvie1974,Feldman1978,Ogilvie1989,Aellig2001,McIntosh2011a,Kasper2007,Kasper2012,Zerbo2015,Alterman2019}.
In particular, \ahe's response lags changes in SSN \citep{Feldman1978} and this lag monotonically increases with \vsw\ \citep{Alterman2019}.

Brightpoints (BPs) are localized enhancements at extreme ultraviolet (EUV) \citep{McIntosh2007a}, x-ray \citep{Vaiana1973}, or both wavelengths.
BPs are a key solar activity indicator and, ``represent a set of low-corona small-scale loops with enhanced emission in the extreme-ultraviolet and X-ray spectrum that connect magnetic flux concentrations of opposite polarities.'' \citep{Madjarska2019}
\citet{McIntosh2007a} determined that EUV BPs are likely rooted at the vertices of supergranule cells and the flow of the supergranules in which they are anchored drives them.
As BPs are signatures of solar activity, it is perhaps unsurprising that their occurrence follows the butterfly pattern \citep{McIntosh2014,Leamon2020}, which Sp\"orer's Law \citep{Maunder1903} associates with sunspots \citep{Carrington1863}.

In this Article, we continue the process of connecting solar wind \ab[\he] to the solar activity cycle.
\sect{data} describes our data sources and selection.
\sect{herald} extends observations of \ahe's variation with solar cycle to cover a 45 year period from 1974 until present day.
Here, we present two as-yet undiscussed observations.
\begin{enumerate}
\item Immediately prior to Solar Minima, \ahe\ rapidly depletes and then recovers over \days[\lesssim 250] at all solar wind speeds. We refer to this as \ahe\ or Helium Shutoff.
%\item On the larger scale of its solar cycle variation, \ahe\ has already started to climb , indicating the rising phase of solar cycle 25 has begun.
\item The time between these rapid \ahe\ shutoffs is approximately the same as the time between successive SSN minima.
\end{enumerate}
\sect{bp} contextualizes \ahe\ Shutoffs prior to Solar Minima with EUV BPs.
This connection relates \ahe's rapid depletions or shutoffs with the time period when two solar cycles overlap and the equatorial bands of the Sun's toroidal field related to the older, decaying extended cycle cancel.
\sect{discussion} discusses our results. %\verify{and shows that, based on the time by which \ahe's depletion has preceded Solar Minima 21 through 25, Solar Minimum may have occurred before Dec 2019}.
\sect{conclusion} briefly concludes.
% !TEX root =	../manuscript.tex

\section{Data Sources and Selection} 
\label{sec:data}

This study combines in situ OMNI plasma measurements with \emph{Solar Dynamics Observatory} (SDO) and \emph{Solar and Heliospheric Observatory} (SOHO) remote observations.
Several solar activity indices provide context.

The OMNI database\footnote{\url{https://omniweb.gsfc.nasa.gov/}} contains solar wind magnetic field, thermal plasma, and energetic proton measurements at multiple high\footnote{5 minute and 1 minute (\url{https://omniweb.gsfc.nasa.gov/html/omni_min_data.html})} and low\footnote{1 hour and longer (\url{https://omniweb.sci.gsfc.nasa.gov/html/ow_data.html})} time resolutions \citep{OMNI}.
These measurements are collected from multiple spacecraft, both near-Earth and at the \nst{1} Lagrange point (L1).
This study uses near-Earth Low Resolution OMNI (OMNI/Lo) data.
The L1-collected data is propagated to, ``expected magnetosphere-arrival times,'' (\url{https://omniweb.gsfc.nasa.gov/html/ow_data.html}) and then averaged to 1 hour cadence.
While OMNI/Lo data begins in 1963 and extends to the present day, we limit this study to data collected beginning in 1974.

The \emph{Wind} Solar Wind Experiment (SWE) Faraday Cups (FCs) are the primary source of present day OMNI/Lo data.\footnote{\url{https://omniweb.gsfc.nasa.gov/html/ow_data.html}}
Multiple SWE/FC data products are available \citep{Ogilvie1995a,Kasper2006,Maruca2013a,Alterman2018}.
Per documentation\footnote{Ibid.}, all data are cross-normalized to the SWE/FC data derived according to \citet{Kasper2006}.
Since this study focuses on the steady state solar wind and not transients such as coronal mass ejections (CMEs), we require that OMNI/Lo data satisfy $\ab[\he] \leq 15\%$ and $\vsw < \kms[1000]$.
Prior work has shown that more detailed removal of transients such as CMEs do not significantly change the average \ahe\ \citep{Kasper2007}.
We access the OMNI data using FTP protocols via the \emph{heliopy} \citep{HelioPyV0.11.1} software library.

The \emph{Solar and Heliospheric Observatory} (SOHO) \emph{Extreme-ultraviolet Imaging Telescope} (EIT) \citep{SOHO/EIT} and \emph{Solar Dynamics Observatory} (SDO) \emph{Atmsopheric Imaging Assembly} (AIA) \citep{SDO/AIA} telescopes provide our EUV measurements at 195\AA\ and 193\AA, respectively.
Following \citet{McIntosh2014}, we identify BPs in a manner that accounts for differences in the two instruments.
We make no distinction between quiet Sun and active region (AR) BPs; calculate a daily average of those lying along the central meridian; and then average the daily BP measurements down to 27-day cadence.

We use three solar activity indicators to provide solar cycle context.
The Solar Information Data Center \citep[SIDC]{SIDC,Vanlommel2005} provides our SSN data.
LASP's Interactive Solar Irradiance Data Center \citep[LISIRD]{LISIRD} provides F10.7 cm radio emission and Lyman-$\alpha$ (\Lalpha) data, which demonstrate a response that lags Solar Minima.

% !TEX root =	../manuscript.tex

\section{Heralding the Solar Cycle} 
\label{sec:herald}

\plotSWEAbundance

\cref{fig:SWE-abundance} Panel (A, top) plots the OMNI/Lo \ahe\ as a function of \vsw\ and time.
\vsw\ has been split into 12 quantiles over the entire mission and \ab[\he] within each quantile is averaged down to 250-day time resolution.
Following \citet{Alterman2019} and \citet{Kasper2007,Kasper2012}, the slowest quantile is at the edge of any given instruments operational capabilities and the fastest covers several hundred \kms.
As such, these two quantiles are excluded and \fig{SWE-abundance} covers the \vsw\ range \kms[319] to \kms[602], i.e.\ slow and intermediate speed solar wind.
The legend indicates the middle of each quantile in \kms.
Error bars indicate the standard error of the mean.
Starting in $\sim 1985$, each error bar is smaller than the corresponding marker.

\fig{SWE-abundance} provides solar cycle context with three activity indices.
The 13 month smoothed SSN ($\bm{\DashedLine}$) is plotted against Panel (A)'s secondary y-axis.
By convention, each numbered solar cycle is defined from one Sunspot Minimum to the following Minimum. 
Purple dash-dotted lines (\textcolor[HTML]{ba55d3}{$\DashDotDotDotLine$}) identify established Solar Minima\footnote{\href{http://sidc.oma.be/silso/DATA/Cycles/TableCyclesMiMa.txt}{sidc.oma.be/silso/DATA/Cycles/TableCyclesMiMa.txt}} and the recently announced Solar Minimum 25 during December 2019\MinFootnotes, each labeled in a gray bar at Panel (B)'s bottom.
Panel (B, bottom) plots \Lalpha\ (left, solid blue, \textcolor[HTML]{1f77b4}{$\SolidLine$}) and F10.7 cm radio emission (right, dashed orange, \textcolor[HTML]{ff7f0e}{$\DashedLine$}).
To match SSN data, both have been averaged to monthly cadence and then smoothed with a centered 13 month window.
%For Minimum 25, we use the NASA/NOAA joint prediction\footnote{\url{https://www.swpc.noaa.gov/news/solar-cycle-25-forecast-update}.} of April, 2020.
%The blue band indicates the Minimum 25 $\pm$ 6 month uncertainty.
Visual inspection shows that both \Lalpha\ and F10.7 reach a minimum after SSN.

As \citet{Alterman2019} observe directly with \emph{Wind}/SWE data, OMNI/Lo \ahe\ reached a consistent maximum during cycles 23 and 24 of $4\% \lesssim \ab[\he] \lesssim 5\%$ across \vsw\ quantiles.
Cycle 24's Minimum and declining phase also indicate that the helium abundance has reached a similar value bottoming out at $\sim \! 1\%$ in the slowest speeds.
Extrema 21 and 22 show similar, mutually consistent behavior.
\ab[\he] during these Minima bottom out at $\sim 2\%$ and these maxima peak between $5\% \lesssim \ab[\he] \lesssim 6\%$.
Minimum 23 bottoms out at a value intermediate between the prior and following Solar Minima.
\ab[\he] may also carry signatures of the double peak or Gnevyshev gap \citep[double SSN peaks]{Hathaway2015} across all Sunspot Maxima, which is the study of future work.

\citet{Alterman2019} observed a decline in \ab[\he] during cycle 24's trailing edge.
\fig{SWE-abundance} shows that--on this 250-day timescale--\ahe\ has reached a local minimum and is now ascending towards is Solar Maximum values across all but the slowest \vsw-quantile.
\fig{SWE-abundance} also indicates an as-yet undiscussed feature across multiple Solar Minima:
on the 250-day timescale used here, \textbf{\emph{a sharp departure from \ahe's local sinusoidal trend appears in all bu the slowest one or two \vsw-quantiles before Solar Minima 22 through 25}}.
Vertical dotted lines (\textcolor[HTML]{ba55d3}{$\DotDotDotLine$}) indicate the date of these rapid \ahe\ depletions or shutoffs, averaged across \vsw-quantiles.
The blue bands (\textcolor[HTML]{7fffff}{$\blacksquare$}) on the top and bottom of each panel surrounding the dotted lines are the associated standard deviation of the dates.
%These two features imply that \emph{\textbf{Solar Minimum 25 may have already occurred.}}
These two features demonstrate that \emph{\textbf{\ahe\ shutoff occurs prior to Solar Minima as defined by SSN}}.

\plotAhePreceedingSolarMin

\fig[(\emph{Left})]{Shutoff-dt-SolMin} plots the time by which \ahe\ shutoff heralds the Solar Cycle in orange circles (\textcolor[HTML]{ff7f0e}{$\bm{\Circle}$}).
The uncertainty is calculated in two ways.
Green error bars (\textcolor[HTML]{2ca02c}{\DashedLine}) indicate the uncertainty only due to \ahe's precision ($\mathrm{P}$), i.e.~the standard deviation across \vsw-quantiles that \fig{SWE-abundance} shows as blue bars.
Orange error bars (\textcolor[HTML]{ff7f0e}{\SolidLine}) show the uncertainty on \ahe's heralding time due to the 250-day averaging window length ($\mathrm{W}$) and precision in \ahe\ timing ($\sqrt{\mathrm{P}^2 + \mathrm{W}^2}$).
The (\emph{Right}) panel plots the (\emph{Left})'s uncertainty sources.
Orange diamonds (\textcolor[HTML]{ff7f04}{$\bm{\lozenge}$}) and green squares (\textcolor[HTML]{2ca02c}{$\bm{\square}$}) plot the uncertainties indicated in the (\emph{Left}) panel with the corresponding color.
The red dash-dotted line (\textcolor[HTML]{d62728}{\DashDotLine}) indicates the 250-day averaging window ($\mathrm{W}$).
This method's 250-day averaging window \citep{Alterman2019,Kasper2012,McIntosh2011a,Kasper2007} is the dominant source of uncertainty in determining when \ahe\ shutoff occurs.
Of note, the heralding time precision decreases starting for Solar Cycle 23 when OMNI/Lo begins transitioning to \emph{Wind}/FC data.
%With that caveat in mind, the \ahe\ shutoff typically the Solar Cycle by \months[\lesssim10].

\plotAheCycleLength

\fig{Shutoff-Cycle-Length} compares SSN Cycle length (\textcolor[HTML]{1f77b4}{$\bm{\PlusEmpty}$}) with \ahe\ shutoff cycle length (\textcolor[HTML]{ff7f0e}{$\bm{\Circle}$}), i.e.~the time between consecutive \ahe\ Shutoffs.
SSN cycle length is plotted as a function of the conventionally defined cycle number.
Shutoff cycle length is plotted as a function of the solar cycle during which the end of the Shutoff cycle interval occurs, i.e.~ $N-1$ in comparison to the conventional solar cycle number a given \ahe\ Shutoff heralds.
\ahe\ Shutoff includes the same uncertainties as \fig{Shutoff-dt-SolMin} and, again, the 250-day averaging window length is the dominant source.
Excluding Cycle 24, the \ahe\ shutoff cycle length is nearly identical to the SSN cycle length.

% !TEX root =	../manuscript.tex

\section{Brightpoints and \ahe\ During the Coexistence of Two Solar Cycles}
\label{sec:bp}

\plotAheBP

The magnetic fields underlying the Sunspot and Solar cycles extend beyond 11 years \citep{Cliver2014,Srivastava2018}.
The transition from one solar cycle to the next is not instantaneous and there is a span of time when the toroidal component of the Sun's magnetic field exhibits polarities from both solar cycles.
This is commonly seen in the overlap of adjacent solar cycles in a Butterfly diagram \citep{Maunder1903,McIntosh2014,Leamon2020,Carrington1863} and has contributed to the identification of an extended solar cycle \citep{McIntosh2015,Leroy1983,Legrand1981}.
In this context, Sunspot Minima correspond to the cancelation of oppositely signed magnetic flux from the older of the two overlapping cycles across the solar equator \citep{McIntosh2015}.

\fig{Ahe-BP} zooms in on the time period from 1995 until present day for which SOHO and SDO provide EUV coverage.
Because we are concerned with an event that happens over \days[\lesssim 250] prior to Solar Minimum, this section focuses on Solar Minima 24 and 25.
We show Minimum 23 and the preceding \ahe\ Shutoff for visual reference.
The (\emph{Top}) panel plots \ahe\ as in \fig{SWE-abundance}.
The (\emph{Bottom}) panel plots daily central meridian BPs averaged down to 27 day cadence.
We have applied a $1\sigma$ Gaussian filter\footnote{\href{https://docs.scipy.org/doc/scipy/reference/generated/scipy.ndimage.gaussian_filter.html}{docs.scipy.org/doc/scipy/reference/generated/scipy.ndimage.gaussian\_filter.html}} to enhance its visual clarity.
The color scale is constructed to highlight BP densities $> 0.8$.
Color levels $> 1$ are shown in white, highlighting BPs in the range 0.8 to 1.
In the (\emph{Top}) panel, Solar Minima and \ahe's Shutoffs are plotted as in \fig{SWE-abundance}.
In the (\emph{Bottom}) panel, they are indicated in light blue to contrast with the BP data.

The (\emph{Bottom}) panel also indicates the first 6 PSP near-Sun encounters and Solar Orbiter (SolO) launch in its top right corner.
The gray band (\textcolor[HTML]{707071}{$\SolidBand$}) indicates $launch - \days[70]$ through the exit of encounter 6 $+ \days[70]$.
The yellow bars (\textcolor[HTML]{ffd700}{$\VertLine$}) indicate the time spent below \au[0.25].
The green bar (\textcolor[HTML]{00ff7f}{$\SolidBand$}) that starts at PSP's 4\textsuperscript{th} encounter indicates the time period starting with SolO's Feb 2020 launch \citep{Muller2020}.
We omit \ahe's pre-Minimum 25 depletion date standard deviation on the top of the (\emph{Bottom}) panel so as to not obscure the PSP encounter dates.

Comparing \fig{Ahe-BP}'s (Top) and (Bottom) panels, especially with respect to the time periods surrounding Solar Minima, is striking.
Over the 4 to 5 years surrounding Solar Maxima, BPs display low levels of activity (BP Density $ \lesssim 0.3$) at middle and high latitudes ($\left| \lambda \right| \gtrsim 25\degree$).
As expected \citep{McIntosh2014}, the midlatitude BPs of the emerging cycle become as or more significant than the decaying cycle prior to Solar Minima and this shift is asymmetric.
For example, BP density $> 0.8$ appears in the northern hemisphere prior to the southern hemisphere during the time period of Minimum 24 and in the southern hemisphere before the northern hemisphere prior to Minimum 25. 
Comparing with \ahe\ in the (Top) panel, it appears as though these asymmetric BP emergences are approximately concurrent with the depletion phase of \ahe\ Shutoff to within approximately one 250-day \ahe\ data point.
The small size of the northern hemisphere BP emergence near the start of 2020 is a result of edge effects in the Gaussian filter. 
This particular feature with a BP density $> 0.8$ at $\lambda \gtrsim 25\degree$ extends through to approximately the present day in the unfiltered image.

 % !TEX root =	../manuscript.tex

\section{Discussion}
\label{sec:discussion}

\fig{SWE-abundance} shows that \ahe\ displays the solar cycle variability characteristic of a solar activity index since at least solar cycle 21.
Earlier results extend this trend back to cycles 19 and 20 \citep{Robbins1970,Ogilvie1974,Feldman1978,Aellig2001}.
\fig{SWE-abundance} presents an additional and as yet undiscussed feature of \ahe's solar cycle variation: on this 250-day timescale, \ahe\ rapidly approaches and then recovers from a local minimum that departs from its long term solar cycle trends prior to Solar Minima 21 through 25.
We refer to this feature as the \ahe\ or Helium Shutoff.
%Our measurements also indicate that present day (June 2020, at the time of submission) \ahe\ has already recovered from this Shutoff prior to a notable increase in SSN, \Lalpha, or F10.7.
%
%\deleted{Fig. 1 presents an additional and as yet unreported feature of \ab[\he]'s solar cycle variation: on this 250 day timescale, \ab[\he] rapidly approaches and then recovers from a local minimum that departs from its long term solar cycle trends prior to solar Minima 21 through 25. The complicating factors discussed above do not negate the significance of these rapid depletions because they are all significant with respect to \ahe's solar cycle scale variability and are present in OMNI/Lo data irrespective of the instrument providing the data.}

\MinimaTable
\fig[(\emph{Left})]{Shutoff-dt-SolMin} shows that \ahe\ Shutoff precedes Solar Minima by \months[\sim \! 10].
\fig{Shutoff-Cycle-Length} shows that the time between successive Shutoffs is nearly identical to the Solar Cycle concurrent with the SSN cycle for which the closing \ahe\ Shutoff occurres.
\fig[(\emph{Right})]{Shutoff-dt-SolMin} shows that the largest single source of uncertainty in this analysis is the 250-day averaging windows defined by the now standard methodology we have employed \citep{Alterman2019,McIntosh2011a,Kasper2012,Kasper2007}.

\tbl{ahe-minima} summarizes the statistics in \cref{fig:Shutoff-dt-SolMin,fig:Shutoff-Cycle-Length}.
The average Heralding Time weighted by the inverse Shutoff Precision is \days[229] and weighted by the inverse Shutoff Uncertainty is \days[302].
Comparing the \emph{Heralding Times} in \tbl{ahe-minima}, the time by which \ahe\ Shutoff precedes Solar Minimum 25 is 37\% larger than the next largest heralding time.

\tbl{ahe-minima} also compares the \ahe\ Shutoff and SSN Cycle lengths in units of \days[250] and \years, i.~e.~units of the 250-day averaging window and a more intuitive timescale.
\ahe\ Shutoff cycle 21 and 22 are identical in length to the SSN cycle.
Cycles 23 are the same to within 1 averaging window.
In contrast, \ahe\ Shutoff cycle 24 at least a full averaging window shorter than SSN cycle 24.
While the combined 362-day uncertainty\footnote{$\sqrt{79^2 + 10^2 + 2\times 250^2}$} due to the precision and averaging window length is too large to determine if this one data point is significant, 
%\wc{the 80-day combined precision\footnote{$\sqrt{79^2 + 10^2}$}} 
that all other \ahe\ Shutoffs herald SSN Minima by \days[< 250] and Shutoff 25 has a precision equal to Shutoff 23
suggest that Solar Minimum 25 may have occurred before the Solar Minimum 25's recently announced\MinFootnotes\ date of December 2019.
Future, higher time resolution work examine this in detail.

EUV BPs tend to form at supergranule vertices, also known as g-nodes, where the radial component of the Sun's toroidal field emerges \citep{McIntosh2014}.
They are commonly associated with the motion of giant convective cells \citep{Hathaway2013,McIntosh2014a}.
Their equatorial migration and torsional oscillations \citep{McIntosh2014} along with the magnetic range of influence \citep[MRoI\footnote{MRoI can be understood as a measure of the extent to which the solar magnetic field is open \citep{McIntosh2006}.};][]{McIntosh2014a} with which they are associated indicate that the underlying magnetic fields have deep roots, likely in deep in the convection zone \citep{McIntosh2014a} or tachocline \citep{McIntosh2014}.

Solar magnetic activity often manifests asymmetrically: a new solar cycle typically emerges in one hemisphere before the other \citep{McIntosh2013,McIntosh2014}.
Signatures of a rising cycle will also emerge before the decaying one has completely vanished.
\fig{Ahe-BP} presents \ahe\ and BPs over the time period when the necessary EIT and AIA EUV measurements are available.
It indicates that, to within one 250-day \ahe\ datapoint, \ahe\ Shutoff occurs concurrently with or immediately prior to BP emergence in the leading hemisphere.
During Solar Minimum 24, BPs emerged in the northern hemisphere before the southern.
As expected \citep{Murakozy2016}, the southern hemisphere is now leading as we enter Solar Cycle 25.

\citet{Alterman2019} show that \ahe\ responds to changes in SSN with a phase lag that monotonically increases with \vsw.
Summarizing their work, slow and fast solar wind originate in distinct source regions on the Sun that are associated with distinct magnetic field strengths.
Slower wind is associated with weaker magnitudes; faster wind with stronger fields.
The height at which \ahe\ ionizes is related to the magnetic field's strength.
As such, \ahe's phase lag and the phase lag's \vsw-dependence imply that slow wind emerges from regions that are more sensitive to changes earlier in the solar cycle and these regions are associated with lower altitudes; fast wind is less sensitive to these changes and it emerges from regions associated with higher altitudes \citep{Alterman2019}.
\citeauthor{Alterman2019} suggest that this is a signature of a filtering mechanism that reduces \ahe\ from its photospheric values in a manner that depends on source region.

In contrast, \ahe's rapid pre-Minima depletions and then recoveries--i.e.\ temporary Shutoffs--occur at approximately the same time prior to Solar Minima 24 and 25 for all speeds at least as fast as \kms[349].
As such, a mechanism independent of solar wind source region and distinct from that driving the phase lag drives these Shutoffs.
% and this mechanism is likely fundamentally related to Solar Minimum.
Given that the Shutoffs are concurrent with leading hemisphere BP midlatitude emergence (to within the 250-day averaging window) and BPs from the rising cycle emerge when two adjacent cycles overlap \citep{McIntosh2014}, we infer that these \ahe\ Shutoffs are likely the result of an unique topology of the global solar magnetic field during Solar Minimum, which is driven by the equatorial cancelation of subsurface magnetic flux.

%According to \citet{Russell2019}, ``.. strong toroidal magnetic flux generated near the tachocline is transported toward the surface where a small fraction of it emerges in the form of bipolar active regions that include sunspots. Sunspots are thus a product of the combined internal flux generation and transport processes...''
Under \citeauthor{Parker1955-Sunspots}'s model, the buoyant rise of toroidal magnetic flux to the Sun's photosphere generates sunspots \citep{Parker1955-Sunspots,Charbonneau2005,Cheung2014}.
The equatorward evolution of the Sun's toroidal magnetic field component leads to the Sunspot butterfly diagram \citep{McIntosh2014,Charbonneau2005,Fan2004}.
Following \citet{McIntosh2014,McIntosh2015}, four toroidal magnetic field bands exist during the decaying solar cycle's declining phases.
Each hemisphere contains two bands and adjacent bands have opposite signs.
These bands live for $> \years[11]$ \citep{Cliver2014,Srivastava2018}. The poleward (rising) bands correspond to the younger of the two extended cycles; the equatorial (decaying) bands correspond to the older cycle \citep{McIntosh2014}.
Solar Minima correspond to the cancelation of two older and decaying toroidal field bands at the equator.
If this is the underlying mechanism, then \ahe\ Shutoff could correspond to the annihilation of these two oppositely signed equatorial flux bands and corresponding lack of flux emergence.

The spacecraft providing OMNI/Lo plasma data all have orbits that are either in or close to the Earth-Sun ecliptic plane.
Therefore, they may not be able to measure any helium related to the two poleward bands of the Sun's toroidal field during these short \ahe\ Shutoffs.
As such, we cannot rule out the possibility that \ahe\ Shutoff is due to the ecliptic nature of these spacecrafts' orbits and this signature is not present in truly polar solar wind.
Solar Oribter's out-of-ecliptic measurements \citep{Zouganelis2020a,Muller2020} may help address this.

%
%
%
%
%\subsection{Text to Incorporate}
%
%
%In addition, this study leverages an established 250-day averaging technique to identify \ahe\ Shutoff.
%As Shutoff occurs over a single data point, it likely occurs over a shorter time period.
%Future work will investigate Shutoff duration so that it can be rigorously tied to other solar activity signatures that occurs over faster time scales, including mid-latitude BP emergence.
%
%\dnc{That \ahe\ Shutoff cycle length and solar cycle length are nearly identical is totally expected if they are driven by the the same or related underlying processes.}
%a.
%
%\fig{Shutoff-dt-SolMin} suggests that present and future instrument capabilities --- assuming they continue to improve --- will enable \ahe\ to accurately predict the onset of future solar cycle.
%Future work will apply higher time resolution averaging and smoothing methods to improve the accuracy of \ahe\ Shutoff's predictive ability.
%\cref{fig:Shutoff-dt-SolMin,fig:Shutoff-Cycle-Length} imply \ahe\ Shutoffs may also enable high accuracy determinations of the Solar Cycle's length.
%\ref{fig:Shutoff-Cycle-Length} also suggests that Solar Cycle 25 may have started before the recently announced date.

 % !TEX root =	../manuscript.tex

\section{Conclusion}
\label{sec:conclusion}

We have studied solar wind helium abundance (\ahe) as a function of speed and time over 45 years.
Using OMNI/Lo data averaged down to 250 day time resolution, we have shown that \ahe: 
\begin{enumerate}
\item likely returns to a consistent values at solar cycle extrema during each of the time periods covered by Solar Minima 21 through Maximum 22 and Maximum 23 through Minimum 25; 
\item rapidly depletes and then recovers over a time period no greater than \days[\sim 250] immediately prior to solar Minimum;
\item has recovered from its pre-Solar Minima 25 rapid depletion; and
\item is already increasing across \vsw\ quantiles along its solar cycle scale variability in the present day.
\end{enumerate}
As solar wind from different source regions have different characteristic speeds and \ahe\ at these speeds respond to solar cycle changes with a distinct phase lag \citep{Alterman2019}, the concurrence of \ahe's Shutoffs for speeds at least as fast as \kms[349] implies that \ahe\ Shutoff is unrelated to differences in these solar wind source regions and how they generate the solar wind.
%Given that BPs emerge asymmetrically in the leading and lagging hemispheres as a new solar cycle rises and this asymmetry is tied to the 
That \ahe\ Shutoff is approximately concurrent with the emergence of BPs in the leading hemisphere suggests that \ahe\ Shutoff may be tied to the cancelation of flux from the toroidal component of solar \BB\ during the time period when two adjacent solar cycles overlap, i.e.~Solar Minimum.
In essence, this corresponds to the death of an extended solar cycle.
Therefore, \ahe\ can serve as a solar activity indicator that heralds a new solar cycle's onset before the sunspot record or other activity indicators.
Although our 250-day averaging window introduces significant uncertainty into our determination of \ahe\ Shutoff's time, a comparison of the \ahe\ and SSN cycles suggests that Solar Minimum 25 may have occurred prior to the recently announced December 2019.
%based on the time by which \ahe\ Shutoff has preceded Solar Minima 21 through 24, \tbl{ahe-minima} clearly indicates that \emph{\textbf{Solar Minimum 25 has already occurred, likely in mid to late 2019.}}

PSP \citep{Fox2015} launched in August 2018 and first dropped below \au[0.25] on October \nst{31} of that year.
By the end of 2020, the spacecraft will have made six trips below \au[0.25], the closest coming to within \au[0.2] of the Sun.
SolO launched in February 2020 \citep{Muller2020}.
\fig{Ahe-BP} indicates PSP's first six encounters as small yellow bars along with the time since SolO launched in light green, both in the top-right corner of the bottom panel.
PSP's \nst{1} and \nd{2} near-sun encounters took place during \ahe\ Shutoff.
However, SolO launched after \ahe\ Shutoff recovered.
Based on this figure, we expect PSP--in particular SWEAP \citep{Kasper2017}--to find alpha particles become markedly more prevalent starting in Encounter 4.
We also expect SolO--especially the SWA instrument suite \citep{Owen2020}--to measure an increasingly significant alpha particle abundance that does not suffer from the absence of such measurements early in the mission.

% !TEX root =	../manuscript.tex

\begin{acks}
This is a preprint of an article published in Solar Physics. The final authenticated version is available online at: \href{https://doi.org/10.1007/s11207-021-01801-9}{https://doi.org/10.1007/s11207-021-01801-9}.

The OMNI data were obtained from the GSFC/SPDF OMNIWeb interface at \url{https://omniweb.gsfc.nasa.gov}.
NASA grants 80NSSC18K0986 and NNX17AI18G supported BLA and JCK.
BLA also acknowledges NASA contract NNG10EK25C.
RJL was supported by an award from the NASA Living With a Star program to NASA GSFC.  
SWM is supported by the National Center for Atmospheric Research, which is a major facility sponsored by the National Science Foundation under Cooperative Agreement No. 1852977.

\emph{Wind} celebrated its \nth{25} anniversary in 2019.
The results presented in this paper strongly depend on the spacecraft's long duration observations that span multiple solar cycles.
As such, we wish to extend our gratitude to the teams that have and continue producing data with the instruments and the project scientists who have kept the mission running for the past quarter century.

This project made use of the following software libraries:
IPython \citep{Perez2007}, 
Jupyter \citep{Kluyver2016a}, 
Matplotlib \citep{Hunter2007}, 
Numpy \citep{VanderWalt2011}, 
Pandas \citep{Mckinney2010}, 
Python \citep{Oliphant2007,Millman2011},
HelioPy \citep{HelioPyV0.11.1},
IDL.
\end{acks}

%%%%%%%%%%%%%%%%%%%%%%%%%%%%%%%%%%%%%%%%%%%%%%%%%%%%%%%%%%%%%%%%%%%%%%%%%%%
%% Acknowledgements
%
% \begin{acks}
%
% \end{acks}

%%% %%%%%%%%%%%%%%%%%%%%%%%%%%%%%%%%%%%%%%%%%%%%%%%%%%%%%%%%%%%
%% Bibliography
%
% Using BibTeX
%
\bibliographystyle{spr-mp-sola}
\bibliography{Mendeley.bib}  

\end{article} 
\end{document}